# Reaction at surfaces

## Tetrahedron bonding of C, N, and O involved reaction at surfaces

**Chang Qing Sun**
**Nanyang Technological University**
11/10/2007

E-mail: ecqsun@ntu.edu.sg
URL: www.ntu.edu.sg/home/ecqsun

**Abstract**

In order to gain advanced understanding of the kinetics and dynamics of C, N, and O reacting with a solid surface, it is necessary to consider the reaction from the perspectives of bond formation, bond dissociation, bond relaxation, bond vibration, and the associated charge redistribution and polarization and the energetic response of the involved atoms and valence electrons. The sp-orbital hybridization is found necessary for these concerned reactions associated with strongly anisotropic bonding and valence identities and the localized energy states of bonding pairs, nonbonding lone pairs, and the lone pair induced antibonding dipoles, as well as the hydrogen bond like and C-H bond like states, which could unify the observations using atomistic microscopy, crystallography, electronic spectroscopy, vibronic spectroscopy, and thermal desorption spectroscopy and provide guidelines for materials design.

## I  Introduction

The understanding of the kinetics and mechanisms of the basic elements of C, N, and O reaction bonding to solid surface has been problematic for half a century. Traditionally, chemical reaction at a surface is often considered in terms of hard spheres interacting through electrostatic forces or in terms of adatoms resting in the potential well of adsorption, with little involvement of charge redistribution and polarization. Results probed using diffractional crystallography, atomistic microscopy, electronic spectroscopy, and other techniques are often isolated one from another in analysis. For example, crystallographic and microscopic observations are interpreted in terms of atomic dislocations or surface reconstructions; electronic spectral features are related to the superposition of electronic states of the constituent neutral atoms isolated state.[1,2] A certain adsorbate-induced structure change is often accompanied with numerous explanations arguing about static atomic positions with limited knowledge about the kinetics, dynamics, and consequences of charge redistribution and polarization. A huge volume of freely adjustable parameters often leads to numerous mathematical solutions that need to be certain in physics. Solutions are often referred to the ones optimized with the minimization of the total energy in theoretical calculations or minimization of the R-factor in decoding data of diffractions, with little involvement of charge transportation or polarization that dominate in reality. Understanding the electronic process of reaction and the nature and dynamics of surface bonding, and eventually to grasp with factors controlling bond making and breaking was recognized as the foremost important task for the community of physics and chemistry in the coming ages. [3]

The invention of scanning tunneling microscopy (STM) and spectroscopy (STS) and the combination of the STM/S with other experimental techniques has propelled the progress in the surface reaction. A correlation between the chemical bond, energy band, and the surface potential barrier (BBB) has been developed for the surface reaction of the most basic elements of C, N, and O to surfaces, which has enabled encouraging progress in understanding the surface reaction kinetics.

## II Principle: BBB correlation

The inserted tetrahedra in **Figure 1** illustrate the effect of C, N, and O adsorption to solid surfaces,[4,5] which indicates the essentiality of sp-orbital hybridization for an O, N, and C atom upon reacting with atoms not only in gas, liquid, but also in solid phase, while the latter was ever argued forbidden. The basic tetrahedral structures are adopted from the $CH_4$, $NH_3$, and $H_2O$ molecules by replacing the H atom with an atom of arbitrary element B. The slight deviation of the tetrahedron and the orientation of the tetrahedron are subject to the coordination environment for bonding. The electronegativity of B should be lower than that of C, N, or O. In the process of reaction, holes, non-bonding lone pairs, anti-bonding dipoles, hydrogen bond like and C-H bond like are produced, which add corresponding density-of-states (*DOS*) to the valence band or above of the host surface (Figure 2).[6] Bond forming also alters the sizes and valences of the involved atoms and causes a collective and continual atomic dislocation,[7] which corrugate the morphology or the potential barrier of the surface. The difference between O, N and C is the number of lone pairs, which determines the group symmetry of the basic tetrahedron. The atomistic microscopy and the crystallography mat vary from situation to situation, the basic tetrahedron and the adsorbate derived DOS remain unchanged. The bond geometry and the respective valence *DOS* dominate the performance of a compound solid. For example, dipole formation lowers the work function of the surface while overdosing with the adsorbate creates hydrogen bond like that restores the work function. It is stressed that the nonbonding lone pairs, antibonding dipoles and the hydrogen-like bonds should not be oversighted in both organic and inorganic chemistry and that the tetrahedron exhibits a strongly anisotropic features.

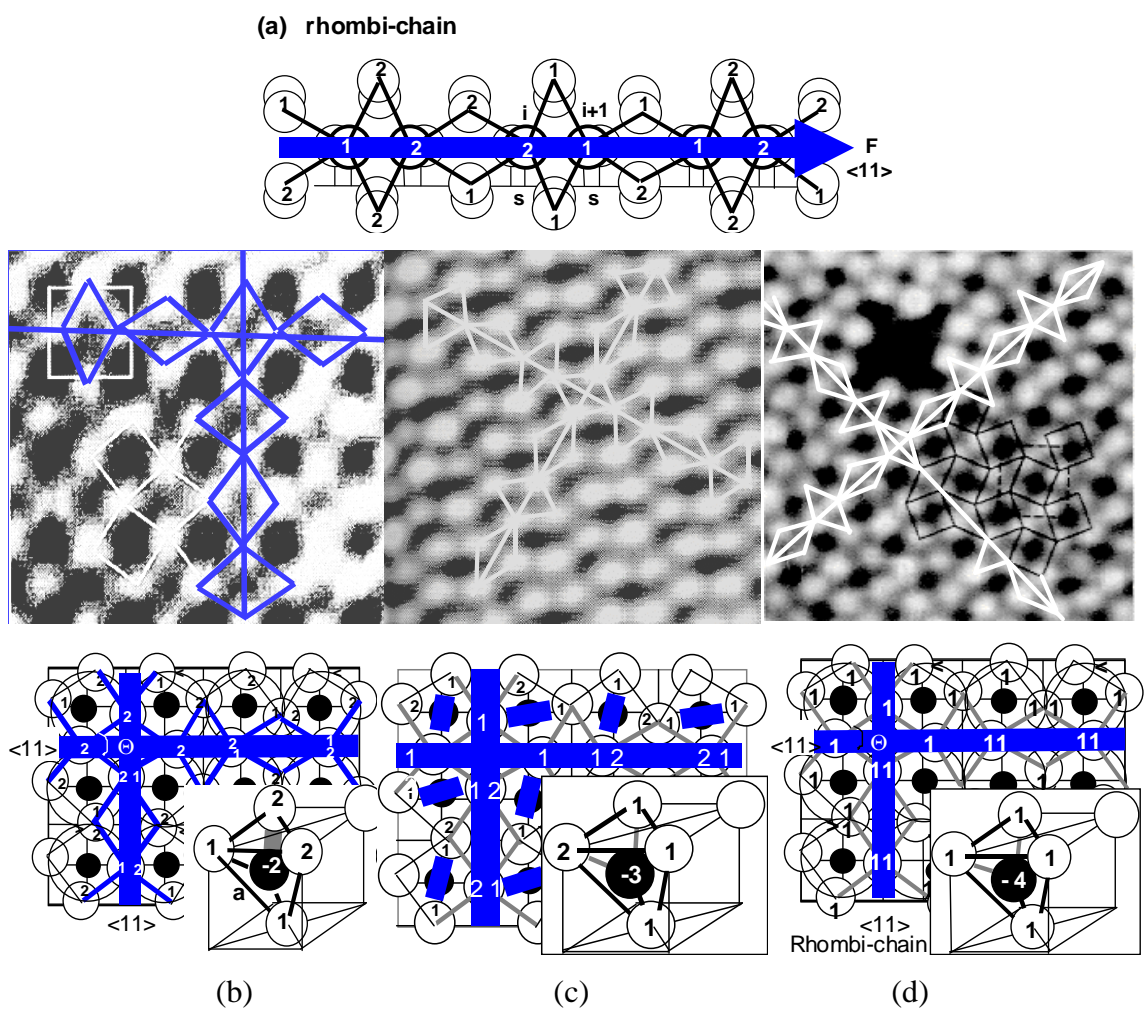

Figure 1 The inserted tetrahedron in (b)-(d) illustrate the basic tetrahedron with labels 1 and 2 standing for different valences induced by $O^{-2}$, $N^{-3}$ and $C^{-4}$ tetrahedron formation: [8]

(b) $B_2O \Rightarrow O^{-2} + 2B^{+1}$ (1) $+ 2B^{dipole}$ (2)

(c) $NB_3 \Rightarrow N^{-3} + 3B^{+1}$ (1) $+ B^{dipole}$ (2)

(d) $CB_4 \Rightarrow C^{-4} + 4B^{+}$ (1).

Formation of the tetrahedron in the surface has led to the (a) non-uniform '- 2 - 1 -- 1 - 2 -' rhombi-chain and the corresponding bond configurations extracted from the STM images of (b) the Rh(001)-(4√2×4√2)R45°-16$O^{-2}$ [9], (c) Ni(001)-(4√2×4√2)R45°-16$N^{-3}$ [10] and, (d) Ni(001)-(4√2×4√2)R45°-16$C^{-4}$ [11].

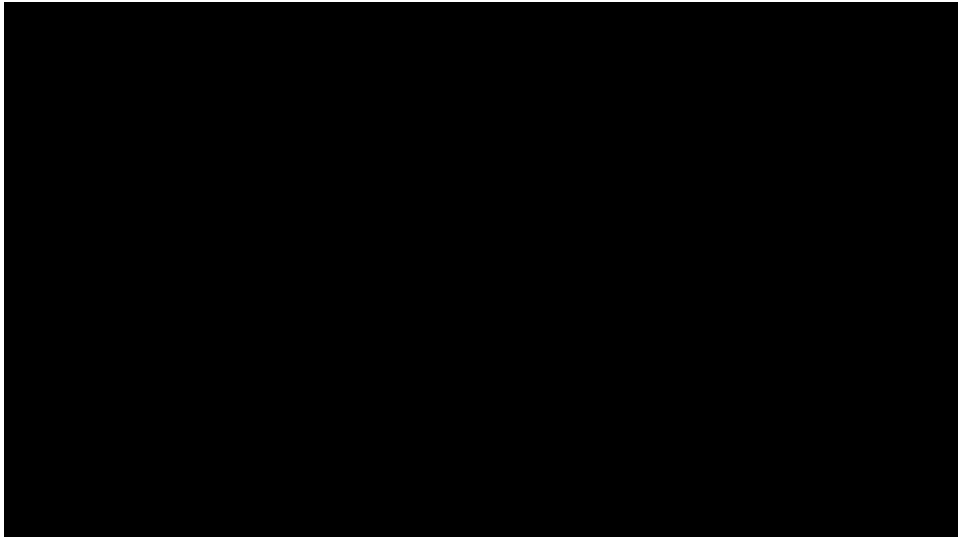

**Figure 2** O and N derived four valence DOS features on metals (upper) and semiconductors (lower):

(1) The $sp^3$ - bonding states that are slightly lower than the 2p-level of the acceptor;

(2) The nonbonding states neither raise nor lower the system energy;

(3) Antibonding (lone-pair-induced dipole) states that reduce the work function from $\phi_0$ to $\phi_1$; and

(4) Hole states under $E_f$ are consequence of electron-transportation in (a) and (b), which changes a conductor to be a semiconductor or widens the band gap of a semiconductor

The DOS feature positions and intensities may vary from situation to situation; the number and their relative positions are intrinsically common. [12] If the lone pairs share the same energy of the electron holes, the corresponding features will not be detectable. For C adsorption, no lone pair feature presents but the C-H like bond induced dipole features remain.

**III Case studies**

3.1 C,N - Ni(001) and O-Rh(001) surface reaction

The reaction kinetics of C, N, and O to the fcc(001) surface shown in **Figure 1** can be expressed as in a B(001)-(4√2×4√2)R45°-16A$^{n-}$ unit cell with labeled atomic valences:

$\Rightarrow 16[O^{-2} + B^{dipole} (2) + B^{+/dipole} (1/2) + B^{+} (1) + Rh \text{ (substrate)}]$

$\Rightarrow 16[N^{-3} + B^{+} (1) + B^{+/dipole} (1/2) + B^{+} (1) + B \text{ (substrate)}]$

$\Rightarrow 16[A^{-4} + B^{+} (1) + B^{2+} (2) + B^{+} (1) + B \text{ (substrate)}]$

Further quantification of the bonding stress from the derived rhombi chain revealed that $N^{3-}$ adsorption induces tensile bonding stress while $C^{4-}$ adsorption derives compressive stress though the STM imaging shows only slight angle difference of derived rhombus.

3.2 Biphase O-Cu(110) reaction kinetics

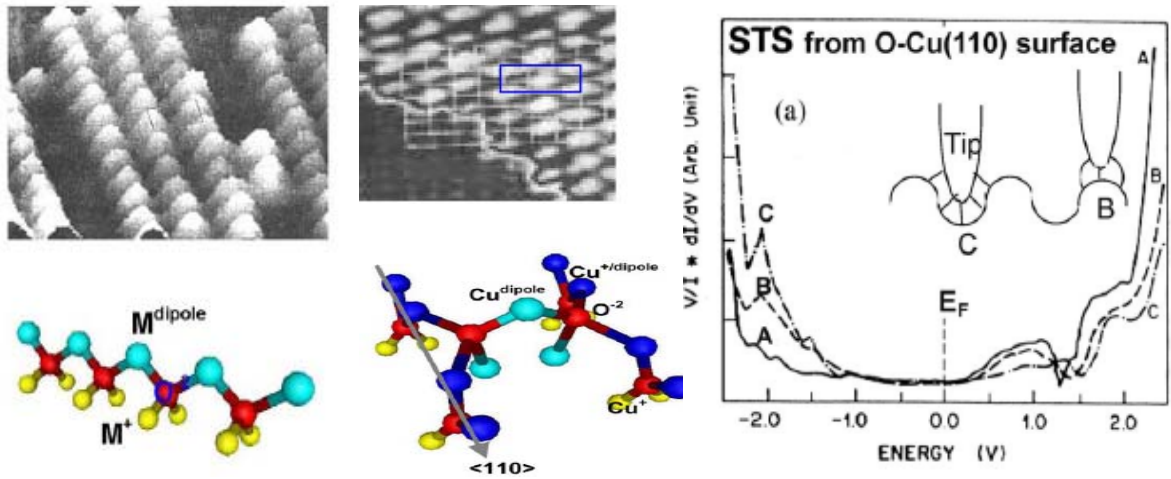

Figure 3 CuO$_4$ tetrahedron built the Cu$^{dipole}$ : O$^{2-}$ : Cu$^{dipole}$ chains on Cu(110)-(2×1)-O$^{2-}$ surface and the Cu$^{dipole}$ : O$^{2-}$ : Cu$^{dipole}$ chains are networked on the Cu(110)-(c(6×2)-8O$^{2-}$ surface. The STS spectra show the DOS of nonbonding lone pair (-2.0 eV) and the lone-pair induced antibonding dipoles (0.5-2.0 eV).[37]

**Figure 3** show the biphase reaction kinetics on the O-Cu(110) surface and the corresponding STM image and STS profiles and the bond configurations. The Cu(110)-(2×1)-O$^{-2}$ surface reaction kinetics is formulated by the following equation with identification of the atomic valence states:

$O_2$ + 4 Cu (surface) + 4 Cu (substrate)

$\Rightarrow$ 2 [$O^{-2}$ + 2$Cu^+$ (substrate) + 2 [Cu (missing-row vacancy) + $Cu^{dipole}$ (Buckled row)]

The second Cu(110)-c(6×2)-8$O^{-2}$ phase reaction kinetics can also be formulated with specification of the atomic valences in different sublayers.[37]

3.3 Four-stage $Cu_3O_2$ binding kinetics on Cu(001) surface

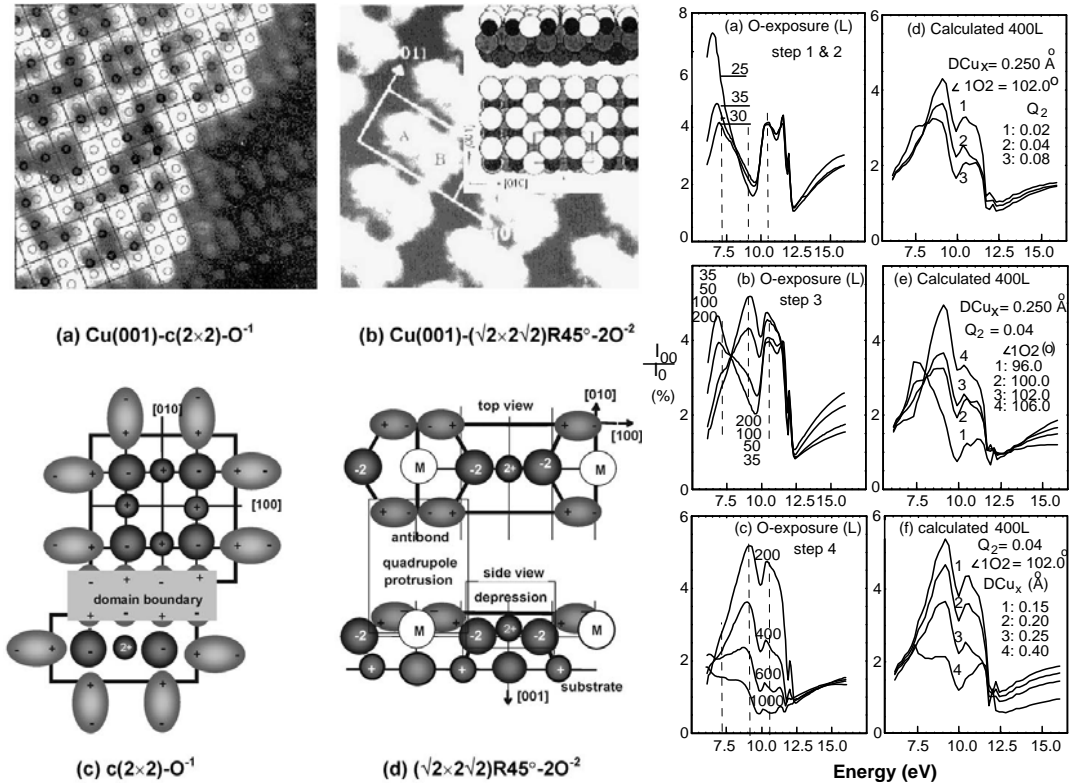

Figure 4 Quantification of the geometry and the four-stage $Cu_3O_2$ bonding kinetics at Cu(001) surface derived from the STM images of the biphase kinetics and the VLEED spectroscopy [13,14]. Multimedia movie is available at doi: 10.1016/S0079-6425(03)00010-0 or at: www.ntu.edu.sg/home/ecqsun/Cu2O3-kinetics.swf

The O-Cu(001) bi-phase ordering arises from the effect of the $O^{-1}$ and subsequently the hybridized-$O^{-2}$ formation. The complete process of reaction is formulated as (**Figure 4**):

$O_2$ (adsorbate) + 4Cu (surface) + 2Cu (substrate)

$\Rightarrow$ 2$O^{-1}$ + $Cu^{2+}$ (surface) + 3$Cu^{dipole}$ ($O^{-1}$- induced) + 2Cu (substrate)

and then,

$\Rightarrow$ 2$O^{-2}$ + $Cu^{2+}$ (surface) + 2$Cu^+$ (substrate) + 2$Cu^{dipole}$ + Cu (M, missing row vacancy)

Based on the STM observations, very-low-energy electron diffraction (VLEED) and XRD calculations, the bond geometry and the four-stage $Cu_3O_2$ bonding kinetics on Cu(001) surface have been quantified. It has been found that the Cu-O bond length is 0.163 (grey-red in **Figure 5**) and 0.176 nm (yellow-red) and the $O^{2-}$ : $Cu^{dipole}$ (red-blue) is 0.195 nm and the angle between the two O-Cu bonds is 104° and the nonbonding angle is 140°, in both the O-Cu(001) and O-Cu(110) surfaces. Actually, the O adsorbate always buckles in the surface to form the tetrahedron instead of resting on the top of the surfaces unless in the $O^{1-}$ state. VLEED calculation reveals clearly four-stage bonding kinetics. It becomes clear that the missing row formation in the second Cu(001)-(√2×2√2)R45°-$2O^{2-}$ phase arises from the saturation of tetrahedron bond formation. The extra Cu atoms in the missing-row are isolated from the bonding, being the same the missing row in the Cu(110)-(2×2) -$2O^{2-}$ phase.

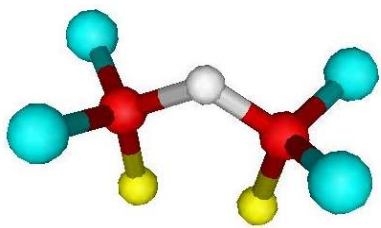

Figure 5 $2O^{2-}$ derived $Cu_3O_2$ pairing tetrahedron on the Cu(001)-(√2×2√2)R45°-$2O^{2-}$ surface showing the strongly localized, anisotropic, and kinetic features With bond length of 0.163 nm, 0.175 nm and nonbonding length of 0.195 nm**:** $2O^{2-}$(red) + $2Cu^+$(yellow) + $Cu^{2+}$ (grey) + $4Cu^{dipole}$.

3.4 O-Co,Ru(10-10) triphase ordering

As shown in **Figure 6**a, the randomly deposited $O^{1-}$ induces metal dipoles. The oval-shaped protrusion in the second Co(10-10)-c(2×4)-$4O^{2-}$ phase (upper image in **Figure 6**b), is induced by the zigzag O-O chains with lacking of Co atoms for the tetrahedron. The lacking of Co atoms for the tetrahedron bonding is compensated by the electron clouds the two dipoles. In the third Co(2×1) -$2O^{2-}$ phase (**Figure 6**c), the observed protrusions correspond to the electron cloud yet the STM dark spots corresponding to the $O^{2-}$ location and Co ions are located in between the $O^{2-}$ ions along the (0001) direction.

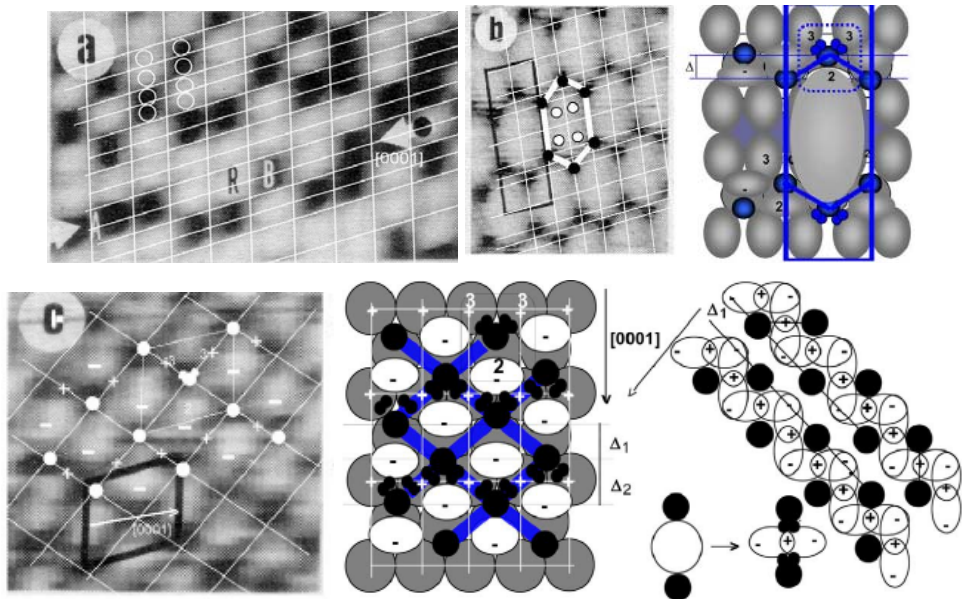

Figure 6 The reaction kinetics of O-Co(Ru)-(10-10) is in the following steps:

Phase I(a): Random located on-top $O^{1-}$ induced patterns.

Phase II(b): In the $c(2\times4)$-$4O^{-2}$ unit cell,

$2O_2$ + 8B (surface) + 8B (substrate) $\Rightarrow$ 4 [$O^{-2}$ + $B^+$ (substrate) + $(2B^{dipole})^+$ + B (substrate)]

Phase III(c): (H-like bond formation):

$O_2$ + 2B (surface) + 2B (substrate) $\Rightarrow$ 2[$O^{-2}$ + $B^+$ (substrate) + $B^{+/dipole}$]

3.5 $C_2H_2B_2$ cluster boding

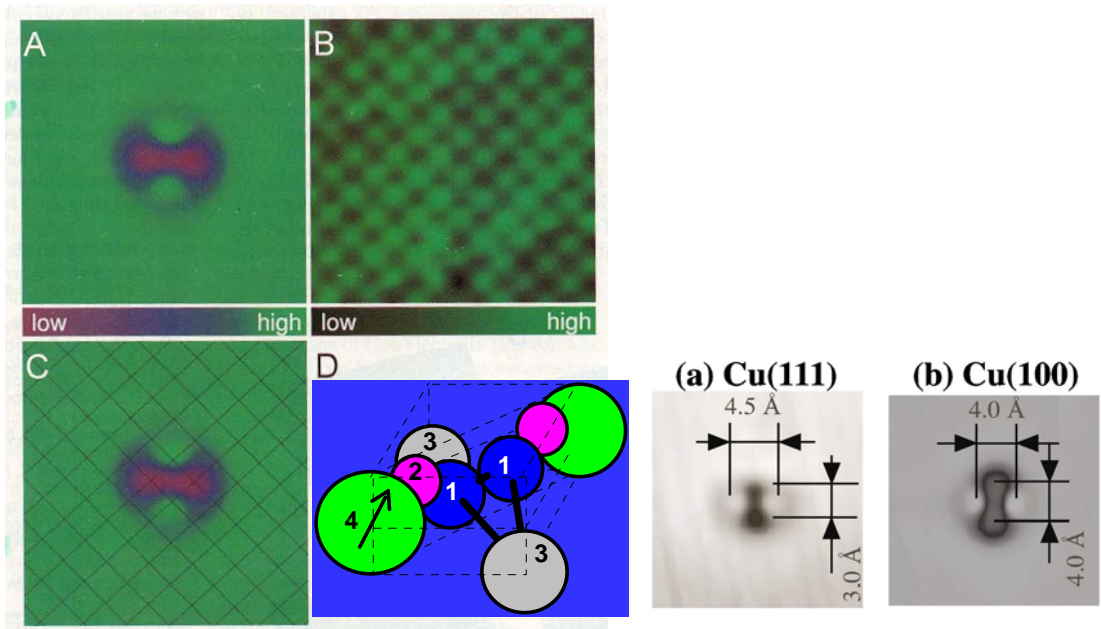

**Figure 7** The concept of tetrahedron formation can also be applied to the $C_2H_2$ molecular adsorption to metal surface. STM images from the $C_2H_2$-Cu(001) [15] and $C_2H_2$-Cu(111)

surface [16] suggested C tetrahedron bond configuration for the reaction kinetics of $C_2H_2M_2$ cluster bonding: $C_2H_2 + 4\,Cu \Rightarrow 2[H^+ + (C\text{-}C)^{-2} + Cu^{+2} + Cu^{dipole}\,(H^+\text{-induced})]$.

3.6 Lone pair vibration

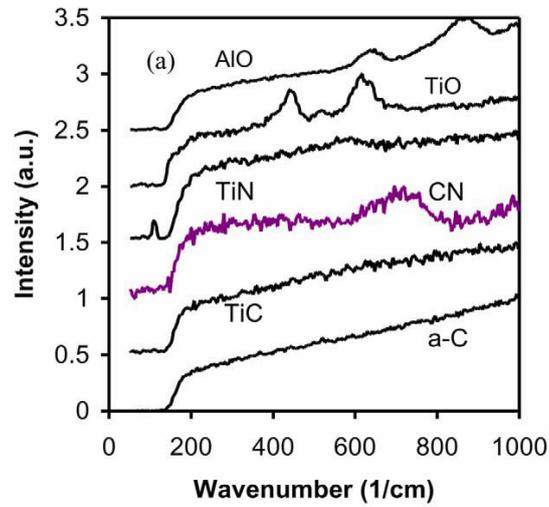

Figure 8 Low-frequency Raman shifts indicate that weak nonbonding interaction (~0.05 eV) exists in Ti and Al oxides and TiN and amorphous CN, which correspond to the non-bonding electron lone pairs generated during the sp-orbital hybridization of nitrogen and oxygen. Therefore, the peak intensities of oxides are stronger than that of nitrides while there are no such peaks at all for a-C and TiC films [17].

3.7 Findings

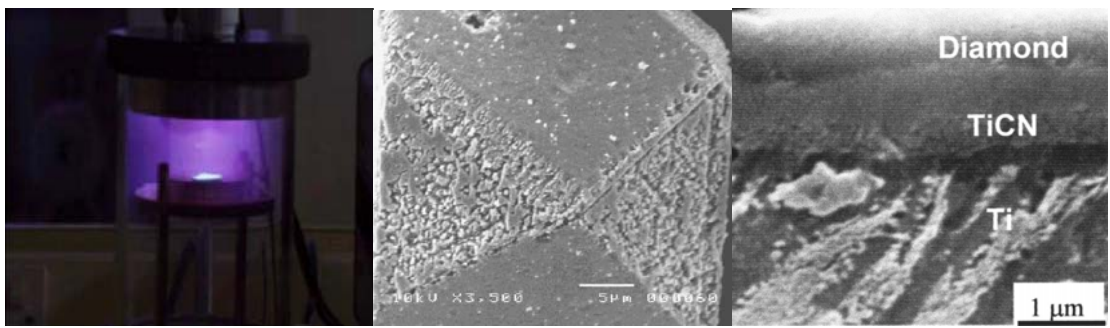

**Figure 9** Knowledge has driven discoveries of PZT blue light emission [38], diamond (111) plane preferential oxidation with the mechanism of the geometrical selectivity of oxide tetrahedron formation [26], and invention of the graded TiCN buffer layer neutralizing interfacial bond stress to improve diamond-metal adhesion [39].

## IV  Summary

The BBB correlation premise and the associated approaches have enabled the following progress:

(i) Generalization of the reaction dynamics and kinetics of over 30 O-derived phases on transition metals such as Cu,[18] Co,[19] Ag and V[20] noble metals such as Rh,[21,22] Ru,[23,24] and Pd,[25] and non-metallic diamond surfaces,[26] and of C/N on Ni(001) surfaces [27,28] as well as $C_2H_2$-Cu(001) surface using formulae of reactions with clear specification of atomic valence evolution and bond forming kinetics at the surfaces. It has been uncovered[29] that O and N act as both accepters by producing holes in the valence band edge and donors by adding the localized lone pair and dipole states around the Fermi level. This achievement forms the hitherto consistent and comprehensive understanding of O, N, and C surface bond formation.

(ii) Unification of the adsorbate-derived signatures of scanning tunneling microscopy/spectroscopy (STM/S), low-energy electron diffraction (LEED), x-ray diffraction (XRD), photoelectron spectroscopy (UPS and XPS), thermal desorption spectroscopy (TDS), electron energy loss spectroscopy (EELS) and Raman spectroscopy, in terms of atomic valence, bond geometry, valence *DOS*, bond strength and bonding kinetics.[30,31,32,33] This practice enhances in turn the capacities of these probing techniques in revealing details of bond forming kinetics and its consequence on the behavior of the involved atoms and valence electronics.[34,35,36]

(iii) Most strikingly, a $Cu_3O_2$ pairing-tetrahedron bond geometry and its four-stage forming kinetics on the O-Cu(001) surface has been quantified using LEED and STM calculations as: one bond forms first and then the other follows; the sp-orbital then hybridizes with creation of lone pairs that induce neighboring dipoles (refer to STM and LEED data in **Figure 3**. The four-stage bonding kinetics has been confirmed to hold general to other analyzed systems as detected using TDS and UPS.[37] Analysis of the O, N, C reaction with the fcc (001) surface of Ni and Rh surfaces has led to an estimation of the bond strain and bond stress, which has driven a design of the TiCN graded buffer layer to neutralize the interfacial stress for diamond deposition on metals.

(iv) It has been uncovered[37] that formation of the basic tetrahedron, and consequently, the four-stage bond forming kinetics and the adsorbate-derived *DOS* features, are *intrinsically* common for all the analyzed systems though the patterns of observations may vary from situation to situation. What differs one surface phase from another in observations are: (a) the site selectivity of the adsorbate, (b) the order of the ionic bond formation and, (c) the orientation of the tetrahedron at the host surfaces. The valences of the adsorbate, the scale and geometrical orientation of the host lattice and the electronegativity of the host elements determine these specific differences *extrinsically*.

New knowledge derived from the BBB correlation has enabled discoveries of new measures or functional materials for blue-light emission (PZT, **Figure 9**, left),[38] diamond preferential oxidation (middle),[26] diamond-metal adhesion with graded TiCN buffer design (right),[39,40] photonic switch (opal array infilled with ferroelectrics),[41,42] electron emission (nitrogenation and stress),[43,44,45,46,47] nitride self-lubrication (CN and TiN),[48] magnetization modulation (Fe nitride),[49,50,51,52,53] gate device (Si/SiO$_2$),[54,55] self-assembly growth,[56,57,58,59,60,61,62] bio sensors (CNT and nano-ZnO),[63,64,65] and mechanical performance (stress modulation),[66,67] as well as consistent insight into the joint effects of chemical passivation and physical miniaturization on the behavior of nanosolids.[68]

### V Prospectus

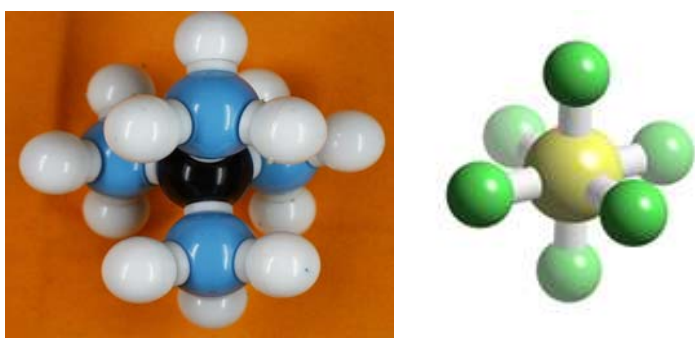

Figure 10  Proposed CF$_4$B$_{12}$ (C$^{4+}$ + 4F$^{1-}$ + 12B$^{dipole}$) and SF$_6$B$_{18}$ (S$^{6+}$ + 6F$^{1-}$ + 18B$^{dipole}$) molecules for actants in syntactic blood. The sp-orbital hybridization of F atom generate three nonbonding lone pairs that induce the B$^{dipole}$.

Progress made insofar has evidence the essentiality of sp-orbital hybridization of C, N, and O involved reaction and the power of the BBB correlation and the associated approaches. With the innovated ways of thinking and the approaches from the perspective of bond formation, bond dissociation, bond relaxation and vibration, the BBB correlation may be extended to reaction of other electronegative elements or molecules such as CF$_4$B$_{12}$ and SF$_6$B$_{18}$ molecules (see **Figure 10** proposed for the actants of syntactic blood), in materials and process design for industrial catalysis, energy conversion, and biosensors, etc. The sp-orbital hybridization may extend to other electronegative elements such as B, P, S, F, Cl, and molecules such as NO, CO, etc. There are indeed plenty of rooms for further exploration. We confident that further synergetic efforts could lead to greater achievement that is even more interesting, useful, and rewarding.

**Acknowledgement**

The practitioner would like to express his sincere gratitude to Professors P.J. Jennings, S.Y. Tong, C.L. Bai, M.F. Ashby, C. Humphreys, S Veprek, J.S. Colligon, A. Stelbovics, S.X. Dou, and E.Y. Jiang, for their encouraging and valuable communications. An encouraging and stimulating discussion in person with Dr Alan G. MacDiarmid at University of Texas at Dallas is particularly gratefully acknowledged and memorized.